\documentclass[10pt,conference]{IEEEtran}

\usepackage{graphicx} % Required for inserting images
\usepackage{booktabs}               % for \toprule etc.
\usepackage{array}                  % (optional) better tables
\usepackage{amsmath,amssymb,amsfonts}
\usepackage{algorithm} % for algorithm environment
\usepackage{algpseudocode} % for pseudocode
\usepackage{subcaption}

\usepackage{textcomp}
\usepackage{xcolor}

\usepackage{tikz}
\usetikzlibrary{shapes,arrows,positioning,fit,backgrounds}
\usepackage{tcolorbox}
\usepackage{todonotes}
\usepackage{enumitem}
\usepackage{siunitx}
\usepackage{multirow}
\usepackage{placeins}
\usepackage{multicol}
\usepackage{xurl}
\usepackage[hidelinks]{hyperref}
\usepackage{xspace}
\usepackage[export]{adjustbox}
\usepackage[numbers,sort&compress]{natbib}
\usepackage{balance}
\newcommand{\qbridge}[0]{\textsc{Q-Bridge}\xspace}

\title{Backend-Aware Graph Learning for Denoising Outcome Distributions in Quantum Program Testing}
\author{%
\IEEEauthorblockN{%
Ning Ma\IEEEauthorrefmark{1},
Jun Dai\IEEEauthorrefmark{2},
Heng Li\IEEEauthorrefmark{1}}%

\IEEEauthorblockA{\IEEEauthorrefmark{1}Department of Computer and Software Engineering, Polytechnique Montreal, Montreal, Canada\\
\{ning.ma, heng.li\}@polymtl.ca}

\IEEEauthorblockA{\IEEEauthorrefmark{2}Mila - Quebec Artificial Intelligence Institute, Montreal, Canada\\
jun.dai@mila.quebec}

}

\begin{document}
\pagestyle{plain}
\maketitle
\begin{abstract}

Testing quantum programs on NISQ (Noisy Intermediate-Scale Quantum) backends is challenging because the noise disturbs outcome distributions and can affect pass/fail decisions. We present \qbridge, a graph learning-based approach that converts noisy observations into denoised distributions suitable for oracle-based verification. \qbridge uses a graph transformer architecture to encode a transpiled quantum circuit, capturing the characteristics of its gates and their connectivity; the physical backend information is encoded together with the logical structure of the circuit. An additional conditioning layer, based on FiLM (Feature-Wise Linear Modulation), takes the encoding as input and integrates noisy observations to produce denoised outcomes. We evaluate \qbridge on 23 IBM noise backends and 6 circuit families representative of practical workloads. In the first setting, we train a separate \qbridge model for each backend; in the second setting, we train a single general model shared across all backends. Across both settings, \qbridge outperforms the state-of-the-art baseline in noise mitigation by a large margin. In testing scenarios with noisy executions, \qbridge achieves 93.97\%--94.90\% precision and 82.50\%--83.51\% recall in detecting bug-induced test failures, significantly outperforming the state-of-the-art baseline. These results indicate that considering the graph structure of the transpiled circuits and the physical characteristics of specific quantum backends is a practical route to more reliable noise-aware quantum program testing.

\end{abstract}


\section{Introduction}
\label{sec:introduction}
Quantum computing has moved beyond physics labs and become publicly accessible through cloud services from major vendors such as IBM Quantum~\cite{steffen2011quantum} and Amazon Braket~\cite{gonzalez2021cloud}, enabling a broader exploration of practical applications~\cite{lau2022nisq}. Despite this progress, current machines remain firmly in the Noisy Intermediate-Scale Quantum (NISQ) era~\cite{li2019tackling}: qubit counts are limited, gate fidelities are imperfect, and environmental disturbances affect computations~\cite{zhong2020quantum}.

As quantum programs scale, effective software engineering practices, particularly testing, become essential~\cite{9474564}. In quantum program testing, correctness is assessed from sampled output distributions rather than deterministic outputs; in particular, in the NISQ setting, device noise can shift these distributions to an extent that alters pass/fail decisions~\cite{ali2023quantum, preskill2018quantum, 10682972}. Prior work adapts classical testing techniques, including combinatorial testing~\cite{9724888}, search-based testing~\cite{10.1145/3510454.3516839}, metamorphic testing~\cite{10172716}, and fuzzing~\cite{wang2018quanfuzzfuzztestingquantum}. However, many evaluations rely on ideal backends that ignore hardware noise, which can yield overly optimistic outcomes and rankings that do not transfer to real devices~\cite{preskill2018quantum}. This makes \textbf{fault--noise discrimination} central: when outputs deviate from the ideal, we must decide whether the deviation arises from a program fault or stochastic device noise~\cite{willsch2018testing}. The problem is compounded by backend variability: a quantum circuit generated by the quantum program must be transpiled to a target backend; the transpilation rewrites gate sets and topology, changing the circuit depth, idle times, and gate counts, thereby altering noise exposure and producing distinct output distributions across backends.

Reliability could, in principle, be ensured through quantum error correction~\cite{terhal2015quantum}, but the qubit overhead of fault-tolerant schemes remains beyond present NISQ capabilities~\cite{RevModPhys.95.045005}. Consequently, quantum error mitigation has become a practical alternative. Mitigation methods reduce bias in measured distributions via algorithmic post-processing while typically avoiding too much auxiliary-qubit overhead~\cite{liao2023machine}. Yet mainstream techniques impose nontrivial costs: Zero-Noise Extrapolation (ZNE) requires repeated executions at multiple effective noise levels~\cite{PhysRevLett.119.180509}, while Clifford Data Regression (CDR) relies on classically simulable near-Clifford training circuits, which becomes increasingly costly as circuit depth and non-Clifford content grow~\cite{Czarnik2021errormitigation}. These limitations motivate learning-based mitigation that infers corrections directly from data.

Recent learning-based approaches are promising, including \texttt{QOIN}, which improves the robustness of distribution-level testing under noise~\cite{10682972}. \texttt{QOIN} trains a separate model per backend and represents executions using summaries of distributions comparing ideal simulations to noisy runs, without explicitly encoding circuit structure, which affects noise-mitigation effectiveness and limits cross-device generalization. Prior work has shown that graph transformers are an effective representation of quantum programs for predicting their fidelity~\cite{wang2022quest} and execution time~\cite{ma2024understandingestimatingexecutiontime}. Moreover, since quantum backends vary in connectivity, calibration, and compiler behavior, we hypothesis that accounting for the physical characteristics of specific backends can help with learning-based noise mitigation. Our observations and hypotheses motivate a graph-aware, backend-conditioned model that leverages circuit structure, shares knowledge across backends, and better supports distribution-level testing based on noisy executions.

To address these gaps, we develop \qbridge, a graph-aware, backend-conditioned denoiser for testing on noisy backends. We organize our study around three research questions.

\begin{itemize}[leftmargin=*,noitemsep,topsep=0pt]
  \item \textbf{RQ1: How accurately can \qbridge mitigate noise on individual noisy backends?}\\
  Prior learning-based mitigation (e.g., QOIN-type methods) relies on distributional discrepancies between noisy and ideal executions, while underutilizing circuit structure and backend-specific factors. In practice, circuit transpilation induces a concrete logical-to-physical mapping and gate schedule, and exposes device-dependent calibration and decoherence effects (e.g., readout error, $T_1$, $T_2$), all of which shape the observed output distribution. We therefore encode the transpiled circuit as a graph and train a graph transformer that integrates topology, calibration and scheduling features, together with shot-derived observation statistics. We evaluate whether this structure- and backend-aware model produces denoised distributions closer to the ideal than methods based only on output summaries.

  \item \textbf{RQ2: Can a single general model mitigate noise across different backends as effectively as per-backend models?}\\
  Training one model per backend is costly and limits reuse. We explore a unified graph transformer that conditions on backend identity, which is pretrained on pooled multi-backend data. We assess whether the general model matches or exceeds per-backend models for noise mitigation across different backends.

  \item \textbf{RQ3: To what extent does \qbridge improve oracle-based quantum program testing?}\\
  The purpose of denoising is to stabilize testing verdicts on noisy hardware, where raw outcomes can confound faults with stochastic noise. We evaluate whether applying \qbridge before oracle checking improves bug detection and reduces misclassification, using both non-failing tests and fault-injected (mutant) programs.
\end{itemize}

{Our work makes several important \textbf{contributions}:

\begin{itemize}[leftmargin=*,noitemsep,topsep=0pt]
  \item \textbf{\qbridge: a graph-aware, backend-conditioned denoiser.}
  We design a Graph Transformer with edge-conditioned attention and a lightweight state-wise conditioning head to correct noisy outcome distributions for testing.

  \item \textbf{Two deployment scenarios.}
  We support both per-backend models and a single general model shared across devices, enabling reuse and cross-backend generalization.

  \item \textbf{Improved distributional accuracy and oracle-based testing.}
  Across diverse backends and circuit families, \qbridge reduces distributional error (Hellinger distance) relative to the state-of-the-art baseline, and when paired with standard test oracles, it improves testing accuracy (higher precision, recall, and F1 score), yielding more reliable test outcomes on noisy backends.
\end{itemize}

% The code and scripts required to reproduce the experiments are publicly available at \url{https://github.com/mooselab/qbridge}.

\textbf{Paper organization.} 
Section~\ref{sec:related-work} discusses previous work related to our study. %Section~\ref{sec:background} introduces the background knowledge of quantum computing related to our work.
Section~\ref{sec:methodology} presents our graph transformer-based denoising model. Section~\ref{sec:setup} describes the setup of the experiments. Section~\ref{sec:result} presents our results for answering the research questions. Section~\ref{sec:discussion} presents further analysis of our model including an ablation study and a scalability analysis.
Section~\ref{sec:threats-to-validity} concerns the threats to the validity of our research. Finally, Section~\ref{sec:conclusion} concludes our study and suggests directions for future research.

\section{Related Work} \label{sec:related-work}

\subsection{Quantum noise mitigation}
Noise in quantum computation arises from environmental disturbances and decoherence, unintended qubit--qubit interactions such as crosstalk, and imperfect calibration, and it accumulates with circuit depth~\cite{10.1145/3464420, joos2013decoherence, ayral2021quantum}. Since full error correction remains impractical on NISQ devices, error mitigation seeks near noise-free estimates without large qubit overheads. Canonical techniques offer different trade-offs. Zero-Noise Extrapolation (ZNE)~\cite{PhysRevLett.119.180509} evaluates a circuit at multiple effective noise levels and extrapolates an ideal value, increasing sampling cost and remaining biased in general. Probabilistic error cancellation yields unbiased estimators with formal guarantees, but its sampling complexity grows rapidly with the desired accuracy. Clifford Data Regression (CDR)~\cite{Czarnik2021errormitigation} learns a regression from noisy to ideal results using near-Clifford surrogates, which limits scalability as non-Clifford content increases and typically targets a single observable.

Learning-based mitigation reduces runtime cost by training predictors that emulate or replace explicit error-mitigation corrections. Under the machine-learning-based quantum error mitigation (ML-QEM) paradigm~\cite{liao2023machine}, mitigation is formulated as supervised prediction from noisy executions to near-ideal estimates; a variety of models~\cite{jayashankar2022quantumerrorcorrectionnoiseadapted} can match classical techniques with fewer shots (number of sample executions) and lower runtime. However, they often rely on paired noisy--ideal references or near-Clifford surrogates, focus on single observables, and are sensitive to backend drift, limiting cross-device generalization. In parallel, noise-aware testing frameworks such as QOIN~\cite{10682972} integrate learned denoising into testing by filtering program outputs before oracle-based assessment, improving verdict stability on hardware. Still, these approaches lack generality in two respects: they typically train separate models for each backend, increasing training cost and limiting cross-backend transfer, and they rely on distribution summaries (instead of circuit-structure-aware representations), which hinders distribution-level reconstruction.

\subsection{Quantum program testing}
Quantum program testing evaluates program correctness under probabilistic semantics on NISQ devices, where noise can flip verdicts~\cite{ramalho2024testingdebuggingquantumprograms, DBLP:journals/corr/abs-2404-06825,bensoussan2025livenoisefingerprintingquantum}. Prior work adapts classical methods for test generation, including combinatorial testing~\cite{9724888}, search-based testing~\cite{10.1145/3510454.3516839}, fuzz testing~\cite{wang2018quanfuzzfuzztestingquantum,10.1145/3763100}, and property-based testing with statistical checks~\cite{10.1145/3428218}.
Assertion styles include statistical, projection-based, and dynamic assertions, and frameworks support unit and integration testing and platform-level differential checks~\cite{10.1145/3307650.3322213, ramalho2024testingdebuggingquantumprograms, DBLP:journals/corr/abs-2404-06825}. 
QuraTest~\cite{10298553} accounts for the fundamental features of quantum programs (magnitude, phase, and entanglement) for automated test generation.
Mutation testing tools (e.g., Muskit~\cite{9678563}, QMutPy~\cite{10.1145/3533767.3543296}) introduce quantum-specific mutation operators and enable large empirical assessments and suite minimization. Metamorphic testing, exemplified by MorphQ~\cite{10172716}, mitigates the oracle problem via relational properties. Differential testing compares executions across tools or backends to surface inconsistencies, with qDiff~\cite{9678792} illustrating compiler/platform-level differential analysis.

\subsection{Quantum circuit representation}
A quantum circuit can be represented as a directed acyclic graph (DAG), where nodes denote gates, qubits, or measurements, and edges encode temporal or data dependencies~\cite{yu2023quantum, tang2022quantum}. Graph Transformers extend attention-based sequence models to graph inputs and capture long-range interactions through layer-wise message passing over the DAG~\cite{vaswani2017attention, velickovic2017graph, yun2019graph}. This representation has been used for tasks such as fidelity estimation by mapping circuit elements to nodes and execution order to edges~\cite{wang2022quest}. Ma et al.~\cite{ma2024understandingestimatingexecutiontime} also leverage a graph transformer with circuit-level context for execution-time prediction. Recent learning-based mitigation methods~\cite{bao2025beyond} also encode circuits as graphs and train transformers to predict mitigated observables, but they often focus on a single expectation value and require Pauli-expectation inputs, increasing measurement cost and complicating transfer across devices. In contrast, we denoise full output distributions using only shot-derived observation features, encode circuit and backend context through edge-biased attention, and enforce a valid probability simplex at inference, aligning naturally with distribution-level testing and enabling cross-backend generalization.

\paragraph*{Positioning and gap}
Existing mitigation methods~\cite{PhysRevLett.119.180509, Czarnik2021errormitigation, 10682972} suffer from several limitations: (a) they target only single observables; (b) they require paired noisy-ideal references or rely on near-Clifford surrogates; (c) they are trained per backend; and (d) they ignore circuit structure beyond shallow summaries. Existing testing frameworks~\cite{9724888,10.1145/3510454.3516839,wang2018quanfuzzfuzztestingquantum} evaluate on ideal simulators or apply learned filtering that is not circuit-aware and does not reconstruct full output distributions, limiting transfer across devices and weakening oracle stability on hardware. While graph-based learning approaches~\cite{wang2022quest,ma2024understandingestimatingexecutiontime,bao2025beyond} show that topology and scheduling carry predictive signal, they have not been coupled to distribution-level denoising for testing.

\qbridge addresses this gap by encoding transpiled circuits as graphs and conditioning attention on backend-specific characteristics via edge-biased scores from calibration and scheduling, enabling distribution-level denoising with shot-derived observation features. We support both backend-wise and general models for improved reuse and cross-backend generalization; integration with standard test oracles yields more reliable verdicts than the state of the art~\cite{10682972}.

\section{Methodology}
\label{sec:methodology}

\subsection{Problem Setup and Notation}
Given a quantum circuit, we transpile it for the target noisy backend to obtain the compiled connectivity and device characteristics, which we refer to as \textbf{graph features}. Executing the transpiled circuit yields measurement counts (with noise), from which we derive \textbf{observation features}. The next subsection (\ref{sec:features}) details the layout of both feature types.

Let $p^{\text{ideal}}_t$ denote the ideal probability of basis state $t$ obtained by executing the circuit on a noise-free simulator. We train a graph transformer model that takes the observation (obtained from a noisy execution) and graph features as inputs and returns a denoised distribution $\hat{p}_t$, so that $\hat{p}_t \;\triangleq\; {p^{\text{ideal}}_t}$.

At inference, the filtered (i.e., denoised) distribution is $\hat{p}_t \propto \text{clip}(\hat{p}_t, 0, 1)$: if the model output $\hat{p}_t$ is less than $0$ or greater than $1$, we set it to $0$ or $1$, respectively. Table~\ref{tab:ideal_noisy_counts} shows the measurement results of a 3-qubit circuit under ideal and noisy execution. In the ideal QASM simulator, all 1024 shots produce the bitstring \texttt{010}, exactly matching the expected output. In contrast, when the same circuit is executed on the noisy backend \texttt{FakeMontreal}, only 932 of the 1024 shots return \texttt{010} (about 91.0\%), while the remaining 92 shots are scattered across other bitstrings due to noise. Our objective is to post-correct the noisy measurement distribution so that it better aligns with the ideal one, thereby making oracle-based test verdicts more stable and reliable. In this case, the probability of state \texttt{010} should be close to $1$ while the others should be close to $0$. For evaluation, we derive observation and graph features from the transpiled circuit and its execution results, feed them to the model to obtain a denoised distribution, and compare the result with the ideal distribution (the oracle).

\begin{table}[t]
\centering
\caption{Ideal and noisy measurement counts for all computational basis states.}
\small
\setlength{\tabcolsep}{3pt}
\begin{tabular}{lcccccccc}
\toprule
 & 000 & 001 & 010 & 011 & 100 & 101 & 110 & 111 \\
\midrule
Ideal result & 0 & 0 & 1024 & 0 & 0 & 0 & 0 & 0 \\
Noisy result & 70 & 1 & 932 & 7 & 1 & 0 & 12 & 1 \\
\bottomrule
\end{tabular}
\label{tab:ideal_noisy_counts}
\end{table}

\subsection{Observation and Graph Features of Quantum Circuits}
\label{sec:features}
\subsubsection{Observation feature}
\label{sec:observation_feature}
Given a quantum circuit executed on a noisy backend with $S$ (e.g., 1024) shots, define
\[
\text{POS}_t \in (0,1),\quad
\text{POF}_t = 1-\text{POS}_t,\quad
\text{ODR}_t = \frac{\text{POS}_t}{1-\text{POS}_t}.
\]
$\text{POS}_t$ is the empirical probability of observing basis state $t$; $\text{POF}_t$ is its complement; $\text{ODR}_t$ is the odds. From Table~\ref{tab:ideal_noisy_counts}, state $t=010$ appears $932$ times out of $1024$ shots, hence $\text{POS}_t=0.910\,(932/1024)$, $\text{POF}_t=0.090\,(92/1024)$, and $\text{ODR}_t=10.130\,(932/92)$.

Finally, for each basis state $t$ we form the \textbf{observation features}
$
x_{\text{obs},t} \;=\; \big[\text{POS}_t,\; \text{POF}_t,\; \log(\text{ODR}_t),\; S^{-1/2}\big] \in \mathbb{R}^{4}.
$

Including both $\mathrm{POS}_t$ and $\mathrm{POF}_t$ provides the model with a complementary view of the empirical outcome for state $t$: the probability mass on $t$ versus the mass on all other outcomes (not-$t$). This makes it easier to decide whether the probability of $t$ should be increased or decreased during post-correction, because any change to $\mathrm{POS}_t$ is immediately reflected in $\mathrm{POF}_t$ (and vice versa). It also helps calibrate the predictor's uncertainty about $t$: values close to $(0.5, 0.5)$ indicate that the measurement outcomes are highly ambiguous between $t$ and not-$t$, whereas values close to $(1, 0)$ or $(0, 1)$ indicate high empirical certainty. Because all outcome probabilities sum to $1$, $\mathrm{POS}_t$ and $\mathrm{POF}_t$ move in opposite directions, making this uncertainty explicit. To further separate likely signal states from noise-induced states, we include $\log(\mathrm{ODR}_t)$~\cite{10682972}. Because $\log(\cdot)$ is undefined at $0$ or $1$, we clip $\mathrm{POS}_t$ to the open interval $(\varepsilon, 1-\varepsilon)$ before forming the odds, using $\varepsilon=10^{-6}$ unless stated otherwise. With finite shots, empirical probabilities can be exactly $0$ or $1$; clipping avoids infinities and improves numerical stability. The term $S^{-1/2}$ encodes the shot-noise scale: for a binomial estimator $\hat p$, $\mathrm{SE}(\hat p)=\sqrt{p(1-p)/S}\le 1/(2\sqrt{S})$, so $S^{-1/2}$ serves as a simple proxy for the measurement uncertainty without access to the unknown ideal $p$~\cite{brown2001interval}.

\subsubsection{Graph feature}
\label{sec:graph_feature}
Each circuit is converted into a directed acyclic graph (DAG), where \textbf{nodes} represent quantum operations (gates) and \textbf{edges} capture the connectivity between them.

\paragraph{Node features.}
For every node (gate) $n$ we form a dense vector $x_n \in \mathbb{R}^{D_{\text{node}}}$ with the exact layout:
\begin{itemize}[leftmargin=*,noitemsep,topsep=0pt]
  \item \emph{Node type one-hot} ($K$ dims): \texttt{in}, \texttt{out}, and all supported gate types. If a node is the first node on a wire (the execution path of the qubit) then $\texttt{in}=1$; if it is the last node then $\texttt{out}=1$. Otherwise it is a gate.  We build a dictionary of all gate types supported by Qiskit: each gate corresponds to an \textit{index}, and the position at \textit{index} + 2 (because the first two positions are \texttt{in} and \texttt{out}) is set to 1, while the others are all 0s.

  \item \emph{Per-qubit calibration (two slots, 4 dims each):} a node can involve one or two qubits, so we construct  two 4-dim slots. For a two-qubit gate, for each involved qubit $q_i$ we store
  $
    \bigl[\tfrac{L_n}{T_1(q_i)},\; \tfrac{L_n}{T_2(q_i)},\; \text{gate\_error}(q_i),\; \text{readout\_error}(q_i)\bigr].
  $
  Here $L_n$ is the gate duration. $T_1$ is the energy relaxation time  and $T_2$ is the dephasing time of the qubit (both from backend calibration). We further include $\text{gate\_error}(q_i)$, the backend-reported average gate error rate for qubit $q_i$, and $\text{readout\_error}(q_i)$, the backend-reported measurement assignment error rate for $q_i$. For single-qubit gates, the second 4-dim slot is zero padded.

  \item \emph{Flag} (1 dim): \texttt{is\_two\_qubit}. The flag is set to 1 for two-qubit gates such as CX, else 0. Two-qubit gates typically have higher error rates due to longer duration and stronger control crosstalk.

  \item \emph{Timing features} (2 dims): the normalized \emph{DAG level} and the normalized \emph{topological position}. The \emph{DAG level} of node $n$ is its earliest parallel layer, defined as the length of the longest dependency path ending at $n$ (thus nodes with the same level can be scheduled in parallel). The \emph{topological position} is the index of $n$ in a fixed topological ordering of the DAG.
  
  \item \emph{Gate parameters} (3 dims): up to three real-valued gate parameters. Each parameter is wrapped \emph{individually} into the interval $(-\pi,\pi]$; missing parameters are set to 0.
\end{itemize}
The final node width is $D_{\text{node}}=K + 14$, where $K$ is the dimension of \emph{Node type one-hot}.

\paragraph{Edge features.}
An edge describes the physical connectivity between two nodes (gates) and can contain the information of execution schedule. We keep edges lightweight and task-focused. For a directed edge $u\!\to\!v$, let
$\mathcal{Q}(n)$ denote the qubits acted on by node $n$ and define the shared-qubit
set $\mathcal{Q}_{uv}=\mathcal{Q}(u)\cap\mathcal{Q}(v)$ (size $\le 2$).
Using the ASAP schedule\footnote{\url{https://quantum.cloud.ibm.com/docs/en/api/qiskit/qiskit.transpiler.passes.ASAPScheduleAnalysis}}, each node has a start time $s(\cdot)$
and duration $L(\cdot)$. For any shared qubit $q\in\mathcal{Q}_{uv}$ the idle
gap from $u$ to $v$ is
$
  \Delta t_{uv}(q) \;=\; \max\!\bigl\{0,\; s(v)-\bigl[s(u)+L(u)\bigr]\bigr\}.
$

We convert the idle gap into \emph{decoherence survival factors}, i.e., multiplicative weights that approximate the probability that the qubit state remains unaffected by relaxation ($T_1$) and dephasing ($T_2$) during the idle period $\Delta t_{uv}(q)$. Formally,
$
  S_{T_k}(q) \;=\; \exp\!\bigl\{-\,\Delta t_{uv}(q)\,/\,T_k(q)\bigr\}\quad(k\in\{1,2\}).
$

Ordering the (at most two) shared qubits as $(q_1,q_2)$ in ascending physical index and
zero-padding the second slot when $|\mathcal{Q}_{uv}|=1$, we define the edge attribute
vector as
$
  e_{uv}
  \;=\;
  \bigl[\,
    S_{T_1}(q_1),\;
    S_{T_2}(q_1),\;
    S_{T_1}(q_2),\;
    S_{T_2}(q_2),\;
    \Delta\mathrm{layer}
  \,\bigr]
  \;\in\;\mathbb{R}^{5},
$

where $\Delta\mathrm{layer}$ is the normalized difference in \emph{DAG levels} between the two nodes. Let $\mathrm{level}(n)$ denote the \emph{DAG level} of node $n$. We define
$
\Delta\mathrm{layer} = \bigl(\mathrm{layer}(v)-\mathrm{layer}(u)\bigr) / \max_{n}\mathrm{layer}(n).
$
This yields a fixed edge feature dimensionality $D_{\text{edge}}=5$. These edge features capture \emph{when} and \emph{how strongly} noise can accumulate along dependency chains. The \emph{decoherence survival factors} encode idle-time decoherence on shared qubits, which directly affects the measured outcome distribution, while $\Delta\mathrm{layer}$ provides a coarse temporal separation signal that helps the model distinguish local gate-to-gate interactions from long-range error accumulation across the circuit depth.

\subsection{Graph Transformer-based Denoising Model}
\label{sec:transformer_model}
We use a graph transformer-based model to denoise the execution outputs of a quantum circuit on a noisy backend. The model consists of two main components: (1) a graph transformer encoder, which encodes the compiled circuit into a graph representation that captures circuit structure and gate interactions; and (2) a FiLM (Feature-wise Linear Modulation) noise head, which conditions this representation on per-state observation features (and backend identity when applicable) to predict a denoised outcome probability for each basis state. We consider two deployment scenarios: (i) backend-wise models trained for a specific device and (ii) a single general model shared across devices to enable cross-backend generalization. The overall model architecture is shown in Fig.~\ref{fig:overview-ours-complete}.
\begin{figure} [h]
    \centering
    \includegraphics[width=1\linewidth]{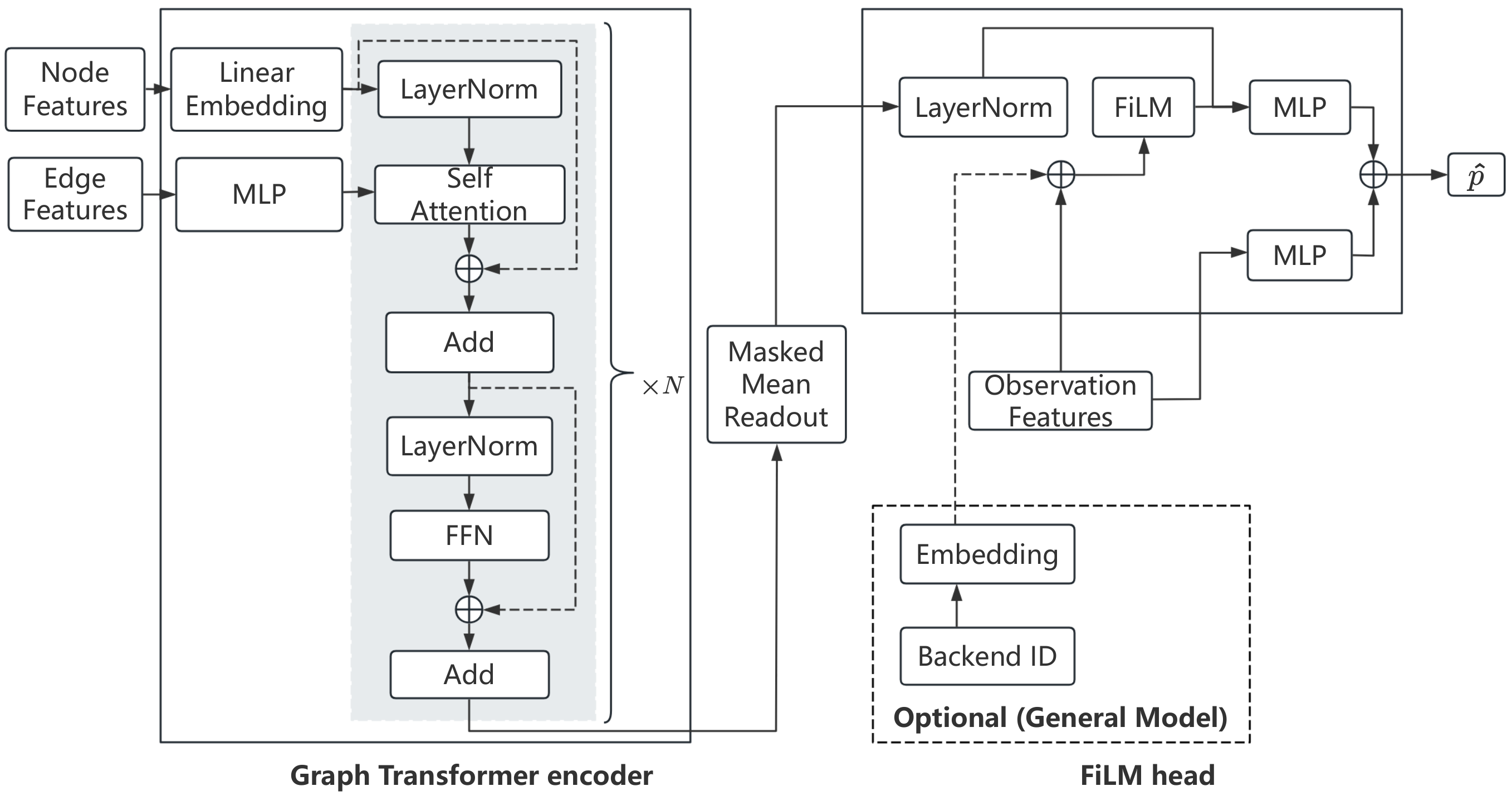}
    \caption{Model architecture overview. Left: Graph Transformer encoder with edge-biased attention. Right: FiLM head that conditions on observation features and an optional learned backend embedding (unified model). Here, MLP denotes a multi-layer perceptron, and FFN denotes the feed-forward network.}
\label{fig:overview-ours-complete}
\end{figure}

\subsubsection{Graph Transformer with Edge-Biased Attention}
\label{sec:encoder}

\paragraph{Purpose, input, and output.}
The encoder maps a variable-size circuit DAG to a fixed-size graph representation $h_{\text{graph}}\in\mathbb{R}^{d}$ for downstream prediction.
Each input graph consists of (1) node features $x\in\mathbb{R}^{N_{node}\times D_{\text{node}}}$, where $N_{node}$ is the number of nodes; (2) directed edges $\text{edge\_index}\in\mathbb{Z}^{2\times E}$, where each column encodes one edge $(u\!\to\!v)$ with the source indices $u$ in the first row and the target indices $v$ in the second row; and (3) edge attributes $\text{edge\_attr}\in\mathbb{R}^{E\times D_{\text{edge}}}$, one feature vector per edge.
With batch size $B$, the encoder returns a batch of graph embeddings $h_{\text{graph}}\in\mathbb{R}^{B\times d}$.

\paragraph{Input embedding and padding.}
We first map each node feature vector to the model width $d$ using a linear projection,
producing token embeddings of shape $X\in\mathbb{R}^{T\times B\times d}$ after batching.
Graphs are padded to the maximum node count $T$ within the batch, and a key padding mask
$\mathbf{M}\in\{0,1\}^{B\times T}$ marks padded positions so that attention
does not attend to them.

\paragraph{Edge-biased self-attention.}
Each encoder layer applies multi-head self-attention with an additive edge bias.
Let $H$ denote the number of heads and $d_h=d/H$ the head dimension.
We compute
$
\text{qkv}=XW_{\text{qkv}},\quad W_{\text{qkv}}\in\mathbb{R}^{d\times 3d},
$
and split $\text{qkv}$ into $Q,K,V\in\mathbb{R}^{T\times B\times d}$, which are then reshaped
into per-head representations (implementation-wise, the $B$ and $H$ dimensions are merged into
$B\!\cdot\!H$).
For edge attributes $\text{edge\_attr}\in\mathbb{R}^{E\times D_{\text{edge}}}$, we produce per-head
scalar biases and write them into a dense bias tensor,
$
\mathbf{B}=\mathrm{DenseBias}\!\big(f_{\theta}(\text{edge\_attr}),\text{edge\_index}\big)\in\mathbb{R}^{B\times H\times T\times T},
$
where non-edge entries are $0$. The attention output is
$
\mathrm{Attn}(Q,K,V)=\mathrm{softmax}\!\Big(\tfrac{QK^\top}{\sqrt{d_h}}+\mathbf{B}\Big)V,
$
with padding positions masked out by $\mathbf{M}$.

\paragraph{Stacking and residual blocks.}
We stack $L$ identical encoder layers, each using pre-norm residual self-attention followed by a
position-wise feed-forward network. Dropout is applied to both attention weights and the feed-forward
block.

\paragraph{Masked mean readout.}
Let $X\in\mathbb{R}^{T\times B\times d}$ be the final-layer token tensor and $\mathbf{M}$ the padding mask.
For each graph $b$, let $V_b=\{t\mid \mathbf{M}_{b,t}=0\}$ denote its valid node positions. We compute a
graph-level embedding by averaging only valid tokens:
$
h_{\text{graph}}^{(b)}=\frac{1}{|V_b|}\sum_{t\in V_b} X_{t,b,:}\in\mathbb{R}^{d},
\quad
h_{\text{graph}}=\big[h_{\text{graph}}^{(1)},\ldots,h_{\text{graph}}^{(B)}\big]^\top\in\mathbb{R}^{B\times d}.
$
This $h_{\text{graph}}$ is the encoder output consumed by the prediction head.

\subsubsection{FiLM Noise Head (State-wise Conditioning)}
\label{sec:film}

\paragraph{Purpose, input, and output.}
The head predicts a denoised probability for each basis state by conditioning the graph embedding on per-state observations. Inputs are a graph embedding $h_{\text{graph}}\in\mathbb{R}^{B\times d}$ and observation features $x_{\text{obs},t}\in\mathbb{R}^{B\times D_{\text{obs}}}$ for each state $t$ as in Section~\ref{sec:observation_feature}. The output is a raw scalar $\hat p_t\in\mathbb{R}^{B}$ per basis state, later clipped to $[0,1]$ at inference.

\paragraph{FiLM conditioning.}
Given the observation vector $x_{\text{obs},t}$ at state $t$, we use a linear projection to produce a
feature-wise scale and shift 
$
[\gamma_t,\beta_t]=W_f\,x_{\text{obs},t},
$
where $\gamma_t,\beta_t\in\mathbb{R}^{d}$ have the same dimension as the graph embedding. Here, $\gamma_t$
controls \emph{how strongly} each embedding dimension is amplified or attenuated, while $\beta_t$ provides
an observation-dependent \emph{offset}. We apply these parameters to a layer-normalized embedding
$
\tilde h_t=(1+\gamma_t)\,\mathrm{LN}(h_{\text{graph}})+\beta_t,
$
where $\mathrm{LN}(\cdot)$ denotes LayerNorm applied over the feature dimension to stabilize training and
make the subsequent modulation less sensitive to embedding scale. A small MLP maps $\tilde h_t$ to a
distribution estimate $\hat p_t^{(1)}$. To ensure that observation features have a direct pathway to the
output, we additionally apply another MLP to $x_{\text{obs},t}$ to obtain $\hat p_t^{(2)}$, and sum the two
branches:
\[
\hat p_t=\hat p_t^{(1)}+\hat p_t^{(2)}.
\]
During fine-tuning, we update only this prediction head and keep the graph encoder fixed.

\paragraph{Backend-wise model.}
In the backend-wise setting, we train a separate model for each device. The observation vector is
$x_{\text{obs},t}\in\mathbb{R}^{4}$ (defined in Section~\ref{sec:observation_feature}), and both the encoder
and the head are trained using data collected from the same backend.

\paragraph{General (cross-backend) model.}
When sharing a single model across heterogeneous devices, backends can differ in global noise patterns
(e.g., overall calibration quality, readout bias, and typical depth inflation after transpilation), even
when the circuit-to-graph construction is identical. To capture such backend-level shifts without
expanding node features, we augment the observation vector with a learned backend embedding
$z_b\in\mathbb{R}^{d_b}$, where $d_b$ is the embedding dimension:
$
x_{\text{obs},t}^{(b)}=\big[\,\text{POS}_t,\;1-\text{POS}_t,\;\log(\text{ODR}_t),\;S^{-1/2},\;z_b\,\big]\in\mathbb{R}^{4+d_b}.
$
We obtain $z_b = E[b]$ from a trainable embedding table $E\in\mathbb{R}^{|\mathcal{B}|\times d_b}$ by indexing the
backend identity $b$, where $\mathcal{B}$ is the set of noisy backends considered in this work ($|\mathcal{B}|=23$ in our
experiments). This embedding allows the FiLM head to condition on backend identity and adapt its mapping
from $(h_{\text{graph}},x_{\text{obs},t})$ to $\hat p_t$, while keeping the encoder shared across devices.

\subsection{Training Objectives}
For a circuit $c$, let $\mathcal{T}_c$ denote the set of measurement outcomes (basis states) considered for that circuit, and let $\hat p_{c,t}$ and $p^{\text{ideal}}_{c,t}$ be the predicted and ideal probabilities of outcome $t\in\mathcal{T}_c$. We first compute a per-circuit mean squared error (MSE) by averaging over its outcomes,
\[
\mathcal{L}(c) \;=\; \frac{1}{|\mathcal{T}_c|}\sum_{t\in\mathcal{T}_c}\big(\hat p_{c,t}-p^{\text{ideal}}_{c,t}\big)^2.
\]
Given a batch $\beta$ of circuits, the training objective is the mean of per-circuit losses:
\[
\mathcal{L} \;=\; \frac{1}{|\beta|}\sum_{c\in\beta}\mathcal{L}(c)
\;=\;
\frac{1}{|\beta|}\sum_{c\in\beta}\frac{1}{|\mathcal{T}_c|}\sum_{t\in\mathcal{T}_c}\big(\hat p_{c,t}-p^{\text{ideal}}_{c,t}\big)^2.
\]
Each circuit may induce multiple training instances, one per measurement outcome $t$ (basis state) paired with its observation features $x_{\text{obs},t}$. Hence, a batch is formed over $(\text{circuit},t)$ instances and may contain fewer distinct circuits than the nominal batch size.

\subsection{Inference and probability clipping.}
At inference, the model outputs a denoised probability $\tilde{p}_t$ for each basis state $t$. To ensure a valid probability value, we clip the output to $[0,1]$:
$
\hat{p}_t=\mathrm{clip}(\tilde{p}_t,0,1)=\max\{0,\,\min\{1,\,\tilde{p}_t\}\}.
$

\begin{table*}[h]
\caption{Hyperparameters of the encoder (edge-biased graph Transformer) and the FiLM head. Pretraining uses $23$ IBM devices with $\geq 7$ qubits ($500$ epochs backend-wise, $1000$ epochs general). Fine-tuning on each CFUT updates only the FiLM head; the encoder is frozen (\texttt{eval}).}
\centering
\small
\setlength{\tabcolsep}{6pt}
\begin{tabular}{lcc}
\toprule
\textbf{Hyperparameter} & \textbf{Pretrain} & \textbf{Finetune (CFUT, head-only)} \\
\midrule
Encoder arch. $(d, H, L, \text{FFN})$ & $(256,\,8,\,4,\,512)$ & \textit{frozen encoder} $(256,\,8,\,4,\,512)$ \\
Dropout / Attn. dropout & $0.1 \;/\; 0.1$ & $0.1 \;/\; 0.1$ (encoder in \texttt{eval}) \\
Optimizer & AdamW & AdamW \\
Learning rate & $5\times10^{-4}$ & head: $1\times10^{-3}$ \quad (optional last layer: $2\times10^{-5}$) \\
Weight decay & $1\times10^{-4}$ & $1\times10^{-4}$ \\
Batch size & 20 & 20 (as available) \\
Epochs & 500 (1000 for general model) & 5--20 (early stop) \\
\bottomrule
\end{tabular}
\label{tab:hyper}
\end{table*}

\section{Experiment Setup} \label{sec:setup}

\subsection{Benchmark Quantum Programs and Selection Criteria}
Following prior practice~\cite{10682972}, we curate 9 benchmark families from two public repositories~\cite{qcuk-code-repo,QuantumAlgorithmZoo} using three criteria~\cite{10682972}:
(i) implementations are written in \textsc{Qiskit} and are publicly available,
(ii) the number of input qubits is at least 3 to ensure a nontrivial input space for circuit generation, and
(iii) the circuits are executable on noisy backends from IBM. In our setting, publicly accessible interfaces in \textsc{Cirq}/Google and in the Rigetti QVM do not expose \(T_1\), \(T_2\), per–gate duration, or readout duration; moreover, the default Rigetti \emph{9q\mbox{-}square} configuration does not provide readout–confusion probabilities or per–gate error rates. Since these quantities are central to our circuit–graph node and edge features, we scope this study to the IBM family.

\subsection{Quantum Execution Backends}
\subsubsection{Noise-free backends}
To obtain noise-free targets, we run the same programs on noise-free simulator (\textsc{Qiskit} Aer, QASM simulator) and collect outcome histograms under the same shot configuration used for noisy runs. These ideal distributions serve as learning targets during training and as reference oracles during evaluation.

\subsubsection{Noisy Backends}
For noisy execution we rely on vendor–supplied noise configurations within native toolchains.
Specifically we use IBM \textsc{Qiskit} Aer with device noise models. Among the 47 available IBM models, we select those with at least 7 qubits to meet benchmark scale requirements, yielding 23 noisy backends in total. The selected 23 IBM backends span diverse sizes, coupler layouts, and calibration profiles~\cite{10682972}, which provide sufficient variety to assess model generalization.

\subsection{Data preparation}
The compilation pipeline is kept identical between ideal and noisy executions in order to isolate the effect of device noise. Each data example consists of a compiled circuit graph with node and edge features derived from topology, scheduling, and calibration, paired with one histogram from a noisy backend and its ideal reference.

\subsection{Training and Evaluation Protocols}

\subsubsection{Pre-training, fine-tuning, and testing data}
We use 9 circuit families in total. Three baseline families (\emph{N CNOT}, \emph{Permutations}, \emph{Expression Evaluation}) are used for pretraining; the remaining six families are CFUTs (Circuit Family Under Test) (\emph{Addition}, \emph{Simon}, \emph{GHZ}, \emph{Binary Similarity}, \emph{Phase Estimation}, \emph{QFT}) and are used for fine-tuning and testing~\cite{10682972}. All circuits are limited to at most \textbf{7} qubits to ensure they can be executed on all $23$ backends. Pretraining pools compiled graphs and paired noisy and ideal distributions from 23 IBM backends. For both the backend-wise and the general models, we use a 90\%/10\% split of the pretraining data for training and validation, where the 10\% validation split is used solely to monitor training, detect overfitting, and select the checkpoint with the lowest validation RMSE, and is not used for final evaluation. Fine-tuning uses predefined tests within each CFUT with the same 90\%/10\% train/validation protocol. Testing uses different inputs from the same CFUT and is strictly disjoint from the training and validation inputs. Because conditional gates activate different branches, different inputs induce different execution paths; consequently, the fine-tuning and testing splits exercise disjoint path combinations~\cite{10682972}.

\subsubsection{Pre-training, fine-tuning, and testing pipeline}
We evaluate two scenarios.

\paragraph{Backend-wise models}
For each backend, we pretrain a transformer baseline, then fine-tune it for each CFUT on a set of predefined tests.
These tests are the sanity checks used during program implementation and are chosen to drive the CFUT through diverse path combinations.
This protocol exposes input-specific and circuit-specific noise characteristics on that backend. The procedure produces 138 fine-tuned models, namely 6 CFUTs over 23 backends.

\paragraph{General model across backends}
We train a single transformer that is shared across devices.
Pretraining uses the pooled data from all 23 backends, and fine-tuning uses the corresponding pooled fine-tuning splits. The procedure yields 6 fine-tuned models, one per CFUT.

\paragraph{Pipeline}
Fig.~\ref{fig:pipeline} shows our workflow. We transpile each circuit to the target backend, extract graph features, run the same compiled circuit on a noise-free simulator for the target distribution, and on noisy backends for observations. We first pretrain the model and then only fine-tune the head of the model per CFUT. At inference time, the model combines the graph features with state-wise observation features to produce a corrected distribution.

\begin{figure}[t]
  \centering
  \includegraphics[width=1\linewidth]{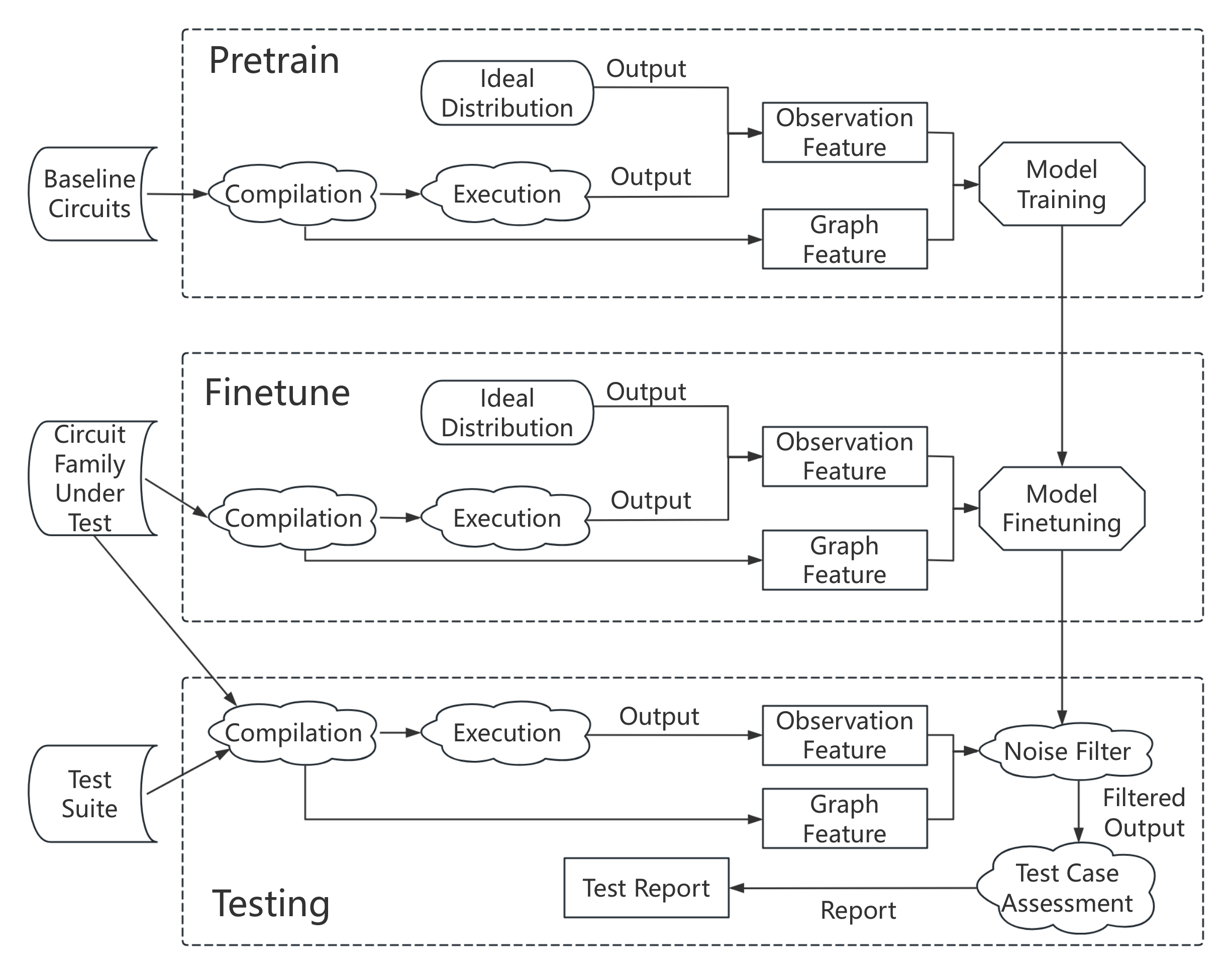}
  \caption{End-to-end workflow used in our study: pretrain on pooled data, fine-tune per CFUT, then testing with UOF/WODF assessment}
  \label{fig:pipeline}
\end{figure}

\subsection{Baselines}
\label{sec:baseline}
We compare \qbridge primarily against \textsc{QOIN}, a state-of-the-art learning-based error mitigation approach, which serves as our main baseline for output-distribution denoising~\cite{10682972}. To provide broader coverage of learning-based mitigation methods, we also include two representative learning-based baselines, \textsc{QRAFT}~\cite{10.1145/3445814.3446743} and \textsc{Q-LEAR}~\cite{10.1145/3510454.3516839}. \textsc{QRAFT} leverages circuit reversibility to infer information about the correct program output through forward-plus-reverse execution. \textsc{Q-LEAR} mitigates noise using a handcrafted feature set defined over circuit-level and output-level characteristics, including depth-cut program error.

\subsection{Evaluation Metrics.}
\label{sec:evaluation_metrics}
Our primary metrics are distributional distances between the predicted (denoised) outcome distribution $\hat{p}$ and the corresponding ideal distribution $p^{\text{ideal}}$. We use the Hellinger distance~\cite{nikulin2001hellinger} as the main metric, as it is symmetric and bounded for probability measures:
\[
H\!\left(\hat p,\,p^{\mathrm{ideal}}\right)
=\frac{1}{\sqrt{2}}\Big\lVert \sqrt{\hat p}-\sqrt{p^{\mathrm{ideal}}}\Big\rVert_2
=\sqrt{\,1-\sum_{t\in\mathcal{T}}\sqrt{\hat p_t\,p^{\mathrm{ideal}}_t}\,}\,,
\]
where $\mathcal{T}$ denotes the set of measurement outcomes (basis states). Hence, $0\le H(\hat p,p^{\mathrm{ideal}})\le 1$: $0$ indicates identical distributions and $1$ indicates disjoint supports.

For test assessment, we report precision, recall, and F1 under the \textbf{Unexpected Output Failure (UOF)}~\cite{9724888, 10.1145/3510454.3516839} and \textbf{Wrong Output Distribution Failure (WODF)}~\cite{9724888, 10.1145/3510454.3516839} oracles.

\subsection{Model Configurations}
Table~\ref{tab:hyper} summarizes the key hyperparameters of \qbridge. Given the number of backends and training scenarios, an exhaustive grid search was impractical. We ran a small pilot sweep and chose a single configuration that worked well across most backends and CFUTs. Unless stated otherwise, we use the same setup for all experiments, with encoder pretraining followed by head-only fine-tuning per CFUT. For pretraining, the backend-wise model runs for 500 epochs and the General model runs for 1000 epochs. For fine-tuning, we update only the head for 5--20 epochs and apply early stopping on validation MSE (minimum 5 epochs, maximum 20 epochs). Each fine-tuned model is trained with 5 different random seeds, and we report the mean and standard deviation of the corresponding test results. All baselines compared with \qbridge follow the same pretraining and fine-tuning protocol, using the same set of random seeds. For \qbridge, pretraining each of the 23 backend-wise models takes 684.36 seconds on average, while pretraining the General model takes 2,174.63 seconds. Fine-tuning each of the 23 backend-wise models takes 22.66 seconds on average, and fine-tuning the General model takes 76.03 seconds.

\section{Results}
\label{sec:result}

\subsection{RQ1: How accurately can \qbridge mitigate noise on individual noisy backends?}
\label{sec:rq1}

\subsubsection{Approach}
We evaluate the performance of \qbridge in the backend-wise scenario and compare it with \textsc{QOIN}, the state-of-the-art baseline. 
As described in Section~\ref{sec:evaluation_metrics}, we use the Hellinger distance $H\in[0,1]$ to measure the discrepancy between the ideal and filtered output distributions, where lower is better. We consider two complementary views. The \emph{backend view} averages $H$ over all CFUTs for each noisy backend. The \emph{circuit view} averages $H$ over all backends for each CFUT. For both views we compare \qbridge with the state-of-the-art baseline \textsc{QOIN} and measure the relative improvement as:
$\boldsymbol{
\mathrm{Relative\ Improvement}=\frac{H_{\textsc{QOIN}}-H_{\textsc{Q-BRIDGE}}}{H_{\textsc{QOIN}}}\times 100\%.}$ To further understand the remaining failure modes, we conduct a backend-level failure analysis by correlating the residual Hellinger distance of \qbridge with calibration-derived backend indicators. Specifically, we compute Spearman correlations between $H_B$ and three indicators: readout-error heterogeneity $\sigma_{\mathrm{ro}}$, average two-qubit gate error $\mu_{2q}$, and two-qubit gate-error heterogeneity $\sigma_{2q}$.

\subsubsection{Results}
Table~\ref{tab:backend-merged} reports the backend-level results on all 23 backends. We focus first on the backend-wise model, shown in the $H_{\text{B}}$ and RI$_{B\leftarrow Q}$ columns. The first 3 columns show the baseline results. Among the baselines, \textsc{QOIN} is consistently the strongest one, with lower $H$ than \textsc{Qraft} and \textsc{Qlear} on every backend. We therefore compute relative improvement with respect to \textsc{QOIN}. Compared with \textsc{QOIN}, the backend-wise model reduces $H$ on all 23 backends, with improvements ranging from 12.1\% on \texttt{FakeParis} to 72.9\% on \texttt{FakeSingapore}. The hardest backends after backend-wise mitigation are \texttt{FakeToronto}, \texttt{FakeRochester}, \texttt{FakeCambridge}, and \texttt{FakeCambridgeV2}, whose residual $H_{\text{B}}$ values remain substantially larger than those of the remaining backends. The last 3 columns of Table~\ref{tab:backend-merged} provide the failure-analysis indicators used to explain these residual errors. They show a clear pattern: larger readout-error heterogeneity, higher average two-qubit gate error, and stronger two-qubit gate-error heterogeneity are all associated with larger residual $H_{\text{B}}$. This is supported by the positive Spearman correlations with $H_{\text{B}}$ ($\rho=0.792$, $0.783$, and $0.737$, respectively). These results suggest that although the backend-wise model effectively exploits schedule-aware structural information, its remaining error is still constrained by backend-specific calibration variability, especially when readout behavior and two-qubit gate quality are highly uneven.

The circuit-level aggregation in Table~\ref{tab:cfut-merged} gives the same conclusion from a different view of the same backend-circuit results. The $H_{\text{B}}$ and RI$_{B\leftarrow Q}$ columns show that the backend-wise model improves all 6 CFUTs over \textsc{QOIN}, with relative gains ranging from 35.8\% on \emph{Simon} to 100.0\% on \emph{GHZ}. The backend-wise model consistently improves over the strongest baseline from both backend-level and circuit-level views.

%\heng{To understand the contribution of the different components of our model (see Fig.~\ref{fig:overview-ours-complete}), we performed an ablation study in which we ...}

\begin{table*}[tb]
\centering
\caption{Backend view. Entries are Hellinger distances $H$ averaged over all CFUTs for each backend (mean $\pm$ sd; lower is better). RI$_{B\leftarrow Q}$ is the relative improvement of the backend-wise model over \textsc{QOIN}, the strongest baseline. RI$_{G\leftarrow Q}$ is the relative improvement of the general model over \textsc{QOIN}. The last 3 columns report selected backend-level indicators used in the failure-mode analysis: readout-error heterogeneity ($\sigma_{\mathrm{ro}}$), average two-qubit gate error ($\mu_{2q}$), and two-qubit gate-error heterogeneity ($\sigma_{2q}$).}
\label{tab:backend-merged}
\scriptsize
\setlength{\tabcolsep}{4pt}
\begin{tabular}{lccccc rr rrr}
\toprule
\multirow{2}{*}{Backend}
& \multicolumn{5}{c}{Average $H$ over CFUTs}
& \multicolumn{2}{c}{Relative Improvement (\%)}
& \multicolumn{3}{c}{Failure-analysis indicators} \\
\cmidrule(lr){2-6}\cmidrule(lr){7-8}\cmidrule(lr){9-11}
& {$H_{\text{Qraft}}$}
& {$H_{\text{Qlear}}$}
& {$H_{\text{\textsc{QOIN}}}$}
& {$H_{\text{B}}$}
& {$H_{\text{G}}$}
& {RI$_{B\leftarrow Q}$}
& {RI$_{G\leftarrow Q}$}
& {$\sigma_{\mathrm{ro}}$}
& {$\mu_{2q}$}
& {$\sigma_{2q}$} \\
\midrule
FakeAlmaden      & $0.693 \pm 0.082$ & $0.178 \pm 0.100$ & $0.080 \pm 0.003$ & $0.030 \pm 0.004$ & $0.043 \pm 0.009$ & 62.5 & 46.3 & 0.048 & 0.024 & 0.017 \\
FakeBoeblingen   & $0.681 \pm 0.077$ & $0.181 \pm 0.100$ & $0.068 \pm 0.002$ & $0.019 \pm 0.002$ & $0.036 \pm 0.010$ & 72.1 & 47.1 & 0.043 & 0.016 & 0.006 \\
FakeBrooklyn     & $0.698 \pm 0.029$ & $0.140 \pm 0.076$ & $0.055 \pm 0.001$ & $0.021 \pm 0.003$ & $0.026 \pm 0.002$ & 61.8 & 52.7 & 0.016 & 0.013 & 0.009 \\
FakeCairo        & $0.694 \pm 0.076$ & $0.103 \pm 0.080$ & $0.041 \pm 0.001$ & $0.023 \pm 0.000$ & $0.024 \pm 0.003$ & 43.9 & 41.5 & 0.020 & 0.013 & 0.011 \\
FakeCambridge    & $0.698 \pm 0.094$ & $0.340 \pm 0.191$ & $0.259 \pm 0.002$ & $0.144 \pm 0.001$ & $0.121 \pm 0.005$ & 44.4 & 53.3 & 0.114 & 0.194 & 0.364 \\
FakeCambridgeV2  & $0.683 \pm 0.074$ & $0.354 \pm 0.209$ & $0.263 \pm 0.003$ & $0.138 \pm 0.009$ & $0.183 \pm 0.004$ & 47.5 & 30.4 & 0.114 & 0.194 & 0.364 \\
FakeCasablanca   & $0.699 \pm 0.077$ & $0.149 \pm 0.100$ & $0.056 \pm 0.001$ & $0.020 \pm 0.003$ & $0.033 \pm 0.000$ & 64.3 & 41.1 & 0.006 & 0.013 & 0.004 \\
FakeGuadalupe    & $0.687 \pm 0.062$ & $0.138 \pm 0.076$ & $0.060 \pm 0.001$ & $0.022 \pm 0.004$ & $0.023 \pm 0.002$ & 63.3 & 61.7 & 0.012 & 0.011 & 0.004 \\
FakeHanoi        & $0.704 \pm 0.062$ & $0.144 \pm 0.087$ & $0.032 \pm 0.003$ & $0.018 \pm 0.001$ & $0.023 \pm 0.004$ & 43.8 & 28.1 & 0.017 & 0.009 & 0.005 \\
FakeJakarta      & $0.727 \pm 0.085$ & $0.143 \pm 0.089$ & $0.067 \pm 0.001$ & $0.028 \pm 0.000$ & $0.025 \pm 0.002$ & 58.2 & 62.7 & 0.010 & 0.013 & 0.005 \\
FakeJohannesburg & $0.719 \pm 0.096$ & $0.246 \pm 0.144$ & $0.110 \pm 0.002$ & $0.049 \pm 0.004$ & $0.058 \pm 0.005$ & 55.5 & 47.3 & 0.039 & 0.028 & 0.010 \\
FakeKolkata      & $0.714 \pm 0.101$ & $0.087 \pm 0.073$ & $0.038 \pm 0.002$ & $0.016 \pm 0.002$ & $0.027 \pm 0.004$ & 57.9 & 28.9 & 0.009 & 0.011 & 0.012 \\
FakeLagos        & $0.695 \pm 0.071$ & $0.138 \pm 0.085$ & $0.040 \pm 0.001$ & $0.013 \pm 0.002$ & $0.015 \pm 0.002$ & 67.5 & 62.5 & 0.003 & 0.008 & 0.003 \\
FakeManhattan    & $0.686 \pm 0.079$ & $0.135 \pm 0.100$ & $0.062 \pm 0.002$ & $0.025 \pm 0.002$ & $0.038 \pm 0.001$ & 59.7 & 38.7 & 0.058 & 0.317 & 0.455 \\
FakeMontreal     & $0.694 \pm 0.077$ & $0.075 \pm 0.080$ & $0.058 \pm 0.001$ & $0.023 \pm 0.005$ & $0.020 \pm 0.002$ & 60.3 & 65.5 & 0.020 & 0.010 & 0.005 \\
FakeMumbai       & $0.694 \pm 0.076$ & $0.098 \pm 0.078$ & $0.045 \pm 0.000$ & $0.021 \pm 0.002$ & $0.029 \pm 0.003$ & 53.3 & 35.6 & 0.021 & 0.010 & 0.007 \\
FakeNairobi      & $0.693 \pm 0.077$ & $0.145 \pm 0.085$ & $0.051 \pm 0.001$ & $0.014 \pm 0.002$ & $0.033 \pm 0.005$ & 72.5 & 35.3 & 0.003 & 0.008 & 0.002 \\
FakeParis        & $0.714 \pm 0.101$ & $0.172 \pm 0.109$ & $0.033 \pm 0.001$ & $0.029 \pm 0.002$ & $0.022 \pm 0.001$ & 12.1 & 33.3 & 0.041 & 0.016 & 0.011 \\
FakeRochester    & $0.703 \pm 0.015$ & $0.471 \pm 0.302$ & $0.384 \pm 0.004$ & $0.191 \pm 0.015$ & $0.175 \pm 0.009$ & 50.3 & 54.4 & 0.097 & 0.124 & 0.256 \\
FakeSingapore    & $0.728 \pm 0.079$ & $0.164 \pm 0.111$ & $0.070 \pm 0.001$ & $0.019 \pm 0.001$ & $0.030 \pm 0.007$ & 72.9 & 57.1 & 0.021 & 0.016 & 0.005 \\
FakeSydney       & $0.685 \pm 0.080$ & $0.123 \pm 0.088$ & $0.057 \pm 0.001$ & $0.024 \pm 0.002$ & $0.026 \pm 0.003$ & 57.9 & 54.4 & 0.027 & 0.018 & 0.027 \\
FakeToronto      & $0.708 \pm 0.096$ & $0.580 \pm 0.291$ & $0.424 \pm 0.003$ & $0.354 \pm 0.006$ & $0.354 \pm 0.008$ & 16.5 & 16.5 & 0.062 & 0.021 & 0.024 \\
FakeWashington   & $0.686 \pm 0.073$ & $0.135 \pm 0.081$ & $0.047 \pm 0.001$ & $0.035 \pm 0.001$ & $0.027 \pm 0.002$ & 25.5 & 42.6 & 0.041 & 0.039 & 0.142 \\
\midrule
Spearman $\rho$ w.r.t.\ $H_B$
& \multicolumn{7}{c}{} & 0.792 & 0.783 & 0.737 \\
\bottomrule
\end{tabular}
\end{table*}

\begin{table*}[tb]
\centering
\caption{Circuit view. Entries are Hellinger distances $H$ averaged over backends for each CFUT (mean $\pm$ sd; lower is better). RI$_{B\leftarrow Q}$ is the relative improvement of the backend-wise model over \textsc{QOIN}, the strongest baseline. RI$_{G\leftarrow Q}$ is the relative improvement of the general model over \textsc{QOIN}.}
\label{tab:cfut-merged}
\scriptsize
\setlength{\tabcolsep}{4pt}
\begin{tabular}{lccccc rr}
\toprule
\multirow{2}{*}{CFUT}
& \multicolumn{5}{c}{Average $H$ over backends}
& \multicolumn{2}{c}{Relative Improvement (\%)} \\
\cmidrule(lr){2-6}\cmidrule(lr){7-8}
& {$H_{\text{Qraft}}$}
& {$H_{\text{Qlear}}$}
& {$H_{\text{\textsc{QOIN}}}$}
& {$H_B$}
& {$H_G$}
& {$RI_{B\leftarrow Q}$}
& {$RI_{G\leftarrow Q}$} \\
\midrule
GHZ               & $0.549 \pm 0.175$ & $0.152 \pm 0.144$ & $0.002 \pm 0.000$ & $0.000 \pm 0.000$ & $0.001 \pm 0.010$ & 100.0 &  50.0 \\
Simon             & $0.586 \pm 0.036$ & $0.168 \pm 0.138$ & $0.053 \pm 0.000$ & $0.034 \pm 0.001$ & $0.034 \pm 0.001$ &  35.8 &  35.8 \\
QFT               & $0.711 \pm 0.031$ & $0.165 \pm 0.152$ & $0.143 \pm 0.001$ & $0.020 \pm 0.000$ & $0.021 \pm 0.002$ &  86.0 &  85.3 \\
Addition          & $0.731 \pm 0.068$ & $0.117 \pm 0.145$ & $0.138 \pm 0.000$ & $0.020 \pm 0.001$ & $0.013 \pm 0.001$ &  85.5 &  90.6 \\
Binary Similarity & $0.725 \pm 0.056$ & $0.248 \pm 0.219$ & $0.217 \pm 0.001$ & $0.089 \pm 0.003$ & $0.092 \pm 0.004$ &  59.0 &  57.6 \\
Phase Estimation  & $0.714 \pm 0.027$ & $0.137 \pm 0.079$ & $0.248 \pm 0.000$ & $0.064 \pm 0.002$ & $0.074 \pm 0.005$ &  74.2 &  70.2 \\
\bottomrule
\end{tabular}
\end{table*}

% \begin{tcolorbox}[colback=black!2,colframe=black!40,title=Takeaways]
% \qbridge is effective at denoising quantum execution outputs, reducing the discrepancy between the ideal distribution and the observed noisy distribution by an average of \textbf{53.6\%} across different backends, compared to the state-of-the-art baseline \textsc{QOIN}. The improvement is consistent across all backends and circuit families.
% \end{tcolorbox}

\subsection{RQ2: Can a single general model mitigate noise across different backends as effectively as per-backend models?}
\label{sec:rq2}

\subsubsection{Approach}
We compare the general model trained across backends with \textsc{QOIN}. Similar to RQ1, the evaluation uses the Hellinger distance $H\in[0,1]$ between the ideal and filtered distributions, with lower values being better.

\subsubsection{Results}
The $H_{\text{G}}$ and RI$_{G\leftarrow Q}$ columns in Tables~\ref{tab:backend-merged} and~\ref{tab:cfut-merged} show that the general model also improves over \textsc{QOIN} on all $23$ backends and all six CFUTs, although its gains are less consistent than those of the backend-wise model. At the backend level, \texttt{FakeCambridge}, \texttt{FakeCambridgeV2}, \texttt{FakeRochester}, and \texttt{FakeToronto} remain the most difficult backends, matching the hard cases found in the backend-wise analysis. The general model is competitive with or better than the backend-wise model on several backends, including \texttt{FakeGuadalupe}, \texttt{FakeJakarta}, \texttt{FakeMontreal}, \texttt{FakeParis}, and \texttt{FakeWashington}; however, the backend-wise model remains stronger on most other backends.

The circuit-level aggregation gives the same conclusion from a different view of the same backend-circuit results. The general model improves all six CFUTs over \textsc{QOIN}; compared with the backend-wise model, it is slightly better on \emph{Addition}, matches it on \emph{Simon}, and is only marginally worse on the remaining CFUTs. Overall, the general model preserves most of the denoising benefit through cross-backend parameter sharing, but backend-specific specialization remains preferable overall. This is consistent with the fact that different backends exhibit distinct and time-varying noise characteristics, which induce distribution shift across devices. A dedicated backend-wise model can adapt directly to backend-specific regularities, whereas the general model must learn a shared representation that remains effective across multiple noise regimes~\cite{Caruana1997MultitaskL}. In addition, the backend embedding acts only as a compact identifier and may not fully capture backend-specific variation; thus, the benefit of pooled training can be partially offset by backend heterogeneity~\cite{ben2010theory}.

% \begin{tcolorbox}[colback=black!2,colframe=black!40,title=Takeaways]
%  \qbridge general model achieves comparable noise-mitigation performance to that of the backend-wise model, outperforming the state-of-the-art baseline by a large margin.
% \end{tcolorbox}

\subsection{RQ3: To what extent does \qbridge improve oracle-based quantum program testing?}
\label{sec:rq3}

According to Table~\ref{tab:backend-merged} and Table~\ref{tab:cfut-merged}, \textsc{QOIN} consistently outperforms \textsc{Qraft} and \textsc{Q-LEAR}. Therefore, in this subsection, we focus on comparing \qbridge only against \textsc{QOIN}.

\subsubsection{Approach}
\paragraph{Test circuits and mutated variants}
To evaluate how \qbridge can improve quantum program testing, we create mutated variants of the original to simulate buggy programs. Ideally, buggy variants should fail the tests, whereas the original circuits should pass them. We use 132 original circuits (from the 6 CFUTs) with 3 mutated variants each; replicated across 23 backends, this yields \(N_{\text{origin}}=132\times23=3036\) bug free circuits and \(N_{\text{mutant}}=396\times23=9108\) variants. Mutant circuits do not automatically produce outputs that differ from the originals. A mutant can still satisfy the specification, so we do not assume all mutants are negatives. To establish ground truth, we run both original and mutant circuits on the noise-free simulator and apply \textbf{UOF} followed by \textbf{WODF} against the specification. Circuits that trigger either oracle are labeled negatives and the rest are labeled positives. Since we do not have access to the original QOIN dataset, we regenerate data using the authors’ released code~\cite{10682972}. Although the total data volume is matched, the class balance differs: our dataset has 9{,}936 positives and 2{,}208 negatives, while the QOIN paper reports 10{,}465 positives and 1{,}679 negatives. All methods in our study are evaluated on the same instances to ensure a fair comparison.

\paragraph{Test assessment.}
We integrate \qbridge with test assessment approaches from prior works~\cite{9724888, 10.1145/3510454.3516839}, yielding two complementary test oracles:
1) \textbf{UOF}:
  Triggers if any observed bitstring key is \emph{not} in the support of the ideal distribution.
2) \textbf{WODF}: 
  Otherwise, triggers if the empirical distribution significantly (over a threshold; we set this threshold to $0.01$, following~\cite{10682972}).

We run the same original and mutant circuits on noisy backends and evaluate four settings: No mitigation, \textsc{QOIN}, \qbridge (backend wise), and \qbridge (general). For each setting we filter the noisy outputs and reapply \textbf{UOF} then \textbf{WODF}, and compute TP, TN, FP, and FN with respect to the simulator based ground truth.

\paragraph{Evaluation of \qbridge and baselines.}
We follow the following steps to evaluate the performance of our approach in detecting buggy variants for each circuit $c$:
\begin{enumerate}[leftmargin=1.25em,noitemsep,topsep=0pt]
  \item Run $c$ on an ideal simulator to obtain the reference distribution $p^{\star}(c)$.
  \item For each noisy backend $b$, run both the original $c$ and its variants on $b$ to obtain noisy distributions $p_b(\cdot)$.
  \item Apply each denoiser, namely No mitigation, QOIN, \qbridge\ (backend-wise), and \qbridge\ (general), to $p_b(\cdot)$ to obtain filtered outputs $\tilde{p}_b^{\mathcal{D}}(\cdot)$.
  \item Evaluate the oracles sequentially on $\tilde{p}_b^{\mathcal{D}}(c)$ against $p^{\star}(c)$:
  \begin{enumerate}
    \item If \textbf{UOF} triggers $\Rightarrow$ label as a negative (failure detected).
    \item Else, test \textbf{WODF}; if it triggers $\Rightarrow$ label as a negative.
    \item Otherwise, label as a positive (no failure detected).
  \end{enumerate}
\end{enumerate}

\subsubsection{Results}

\begin{table*}[!t]
\centering
\caption{Bug-induced test failures detection performance of \qbridge and baselines. Results are reported as mean $\pm$ standard deviation across random seeds.}
\label{tab:test-results}
\begin{tabular}{lrrrrrrr}
\hline
\textbf{Setting} & \textbf{TP} & \textbf{TN} & \textbf{FP} & \textbf{FN} & \textbf{Precision} & \textbf{Recall} & \textbf{F1 Score} \\
\hline
No mitigation                  & 32 & 2{,}208 & 0 & 9{,}904 & 100.00\% & 0.32\% & 0.64\% \\
\textsc{QOIN}                  & $958.8 \pm 16.6$ & $1{,}894.2 \pm 3.0$ & $313.8 \pm 3.0$ & $8{,}977.2 \pm 16.6$ & $75.34 \pm 0.33$\% & $9.65 \pm 0.17$\% & $17.11 \pm 0.27$\% \\
\qbridge (backend\mbox{-}wise) & $8{,}297.7 \pm 18.2$ & $1{,}675.3 \pm 2.1$ & $532.7 \pm 2.1$ & $1{,}638.3 \pm 18.2$ & $93.97 \pm 0.03$\% & $83.51 \pm 0.18$\% & $88.43 \pm 0.11$\% \\
\qbridge (general)             & $8{,}197.7 \pm 18.2$ & $1{,}767.3 \pm 5.1$ & $440.7 \pm 5.1$ & $1{,}738.3 \pm 18.2$ & $94.90 \pm 0.05$\% & $82.50 \pm 0.18$\% & $88.29 \pm 0.09$\% \\
\hline
\end{tabular}

\smallskip\noindent\textit{Total circuits:} \(N=12{,}144\) for all settings.
\end{table*}

The results are reported in Table~\ref{tab:test-results}. The ``No mitigation'' row means that we evaluate UOF/WODF directly on the noisy outputs without any denoising. In this setting, the oracle catches all bug-induced failures (FP=0) but generates massive false alarms, misclassifying 9,904 truly-passing circuits as failing (FN=9,904). This indicates that, without error mitigation, the distributions obtained from the noisy simulator can deviate substantially from the noise-free distributions, making the oracle overly pessimistic and leading it to reject many circuits that should have passed.

For \qbridge, we use the models trained in Sections~\ref{sec:rq1} and~\ref{sec:rq2} for denoising. The reported values are averaged over multiple runs with different random seeds. Compared with \textsc{QOIN}, both variants of \qbridge achieve a much stronger precision and recall balance. Recall increases from $9.65 \pm 0.17$\% for \textsc{QOIN} to $83.51 \pm 0.18$\% and $82.50 \pm 0.18$\% for the backend-wise and general models, respectively. Precision also increases from $75.34 \pm 0.33$\% to $93.97 \pm 0.03$\% and $94.90 \pm 0.05$\%. As a result, the F1 score improves from $17.11 \pm 0.27$\% for \textsc{QOIN} to $88.43 \pm 0.11$\% and $88.29 \pm 0.09$\%.

This improvement is mainly driven by the recovery of many more truly passing cases. The backend-wise model increases TP from $958.8 \pm 16.6$ to $8{,}297.7 \pm 18.2$, while the general model reaches $8{,}197.7 \pm 18.2$. This large TP gain comes with a moderate decrease in TN, from $1{,}894.2 \pm 3.0$ for \textsc{QOIN} to $1{,}675.3 \pm 2.1$ for the backend-wise model and $1{,}767.3 \pm 5.1$ for the general model. Overall, these results show that \qbridge effectively filters noise from the observed output distributions, allowing truly passing circuits to pass the oracle while still detecting bug-induced failures.

% \begin{tcolorbox}[colback=black!2,colframe=black!40,title=Takeaways]
% \qbridge achieves 82.50--83.39\% recall and 93.89--94.85\% precision for detecting bug-induced test failures, improving over the state-of-the-art baseline \textsc{QOIN} by 72.48\%--73.37\% and 59.44\%--60.40\%, respectively. Our results indicate that incorporating circuit and backend context improves the reliability of oracle-based testing on noisy backends.
% \end{tcolorbox}

\section{Discussion}
\label{sec:discussion}
\noindent\textbf{The impact of design decisions.} To understand the contribution of the different components of our model (see Fig.~\ref{fig:overview-ours-complete}), we performed an ablation study in which we changed (add/remove) the components of the original \qbridge architecture one at a time to evaluate their impact.
The ablation study confirms that each architectural component contributes positively to performance. Removing edge-biased attention or FiLM conditioning degrades the performance of both the backend-wise and general models. For the general model, removing the backend embedding causes a substantial increase in $H$, confirming that explicit backend conditioning is important for cross-backend denoising.
Due to the space limit, we include the full ablation study in the replication package. 

%Due to space limits, we include the full ablation study and larger-circuit evaluation in the replication package. We briefly summarize the main findings here.

\noindent\textbf{Generalization to larger circuits.}
We limited the circuit size to 7 qubits since it is the size supported by all 23 evaluated quantum backends. To understand how our findings generalize to larger circuits, we evaluated \qbridge on a larger-circuit set with $8$--$15$ qubits. Since four of the backends used in the main experiments have only 7 qubits, namely \texttt{FakeCasablanca}, \texttt{FakeJakarta}, \texttt{FakeLagos}, and \texttt{FakeNairobi}, this study is restricted to the remaining 19 compatible backends. As expected, the absolute Hellinger distances increase compared with the original small-circuit setting. Averaged over backends, \textsc{QOIN} obtains $H=0.434$, while the backend-wise and general models reduce $H$ to $0.351$ and $0.336$, corresponding to relative improvements of $19.0\%$ and $22.4\%$, respectively. At the circuit-family level, the general model improves over \textsc{QOIN} on all 6 larger-circuit families. These results indicate that larger circuits are more difficult to accurate model, but \qbridge still consistently provides substantial gains over the strongest baseline.
The details are included in the replication package.

\section{Threats to Validity}
\label{sec:threats-to-validity}

\noindent\textbf{External validity}.
We evaluate on vendor noise models in \textsc{Qiskit} Aer for $23$ IBM backends, without real-hardware runs or non-IBM devices. Results may differ on other ecosystems or hardware due to calibration drift, crosstalk, and simulator model mismatch. Evaluating additional vendors and real devices is future work.

\noindent\textbf{Internal validity}
Our main experiments cap circuits at $7$ qubits to support a unified protocol across all $23$ backends, matching the smallest devices and following prior practice~\cite{10682972}. This enables controlled backend-wide comparisons, but limits the extent to which the results generalize to wider and deeper programs. To partially mitigate this threat, we additionally evaluate circuits with $8$--$15$ qubits on the subset of compatible backends. Scaling to tens- or hundreds-qubit programs remains future work.

\noindent\textbf{Construct validity}
Our testing evaluation relies on the UOF and WODF oracle definitions from prior work. These oracles operationalize test failures through distribution-level deviations, but they may not capture all notions of quantum program correctness. In particular, WODF can be sensitive to sampling noise and may produce false positives when the number of shots is limited. We mitigate this threat by using the same oracle definitions and shot budget across all compared methods, so the relative comparison is consistent. Future work should evaluate additional oracle designs and study how oracle choice affects mitigation-driven testing outcomes.

\section{Conclusion}
\label{sec:conclusion}

Testing quantum programs on noisy devices is difficult because noise perturbs outcome distributions and can mislead pass/fail decisions. We presented \qbridge, a graph-aware, backend-conditioned denoiser that combines edge-biased attention with a state-wise FiLM head. Across 23 IBM backends and 6 tested circuit families, \qbridge consistently outperforms the state-of-the-art baseline in noise mitigation, achieving substantially smaller Hellinger distance between the denoised output distributions and the ideal distributions. 
Furthermore, a single general model trained across different backends achieves competitive performance compared to backend-wise models.  
\qbridge, through noise mitigation, improves the reliability of oracle-based quantum program testing. These results show that circuit structure-aware modeling and backend conditioning enable robust cross-device denoising and more reliable test outcomes. Future work will evaluate on real hardware and extend to additional circuit families.

\section{Data Availability}
\label{sec:availability}
The code and scripts required to reproduce the experiments are publicly available at \url{https://github.com/mooselab/qbridge}.

% \FloatBarrier
\balance
\bibliographystyle{ACM-Reference-Format}
\bibliography{q-bridge}
\end{document}